\documentclass[12pt]{article}

\usepackage{epic}
\usepackage{eepic}
\usepackage{graphics}
\usepackage{amsfonts}
  \usepackage{amsthm}
 \usepackage{amsmath}
 \usepackage{amscd}
 \usepackage{amssymb}
 \usepackage{latexsym}

 \numberwithin{equation}{section}
 \newtheorem{thm}{Theorem}
 
 \newtheorem{prop}{Proposition}
 \theoremstyle{definition}

\title{On the integrability of a system describing the stationary solutions in Bose--Fermi mixtures}

\author{Ognyan Christov$^\dag$ and Georgi Georgiev$^\ddag$ \\
$^\dag$ Faculty of Mathematics and Informatics, Sofia University, \\1164 Sofia, Bulgaria\\
$^\ddag$ Department of Mathematics and Informatics, University of
Transport, \\1574, Sofia, Bulgaria}
\date{}

\begin{document}

\maketitle

\begin{abstract}
\noindent We study the integrability of a Hamiltonian system
describing the stationary solutions in Bose--Fermi mixtures in one
dimensional optical lattices. We prove that the system is
integrable only when it is separable. The proof is based on the
Differential Galois approach and Ziglin-Morales-Ramis method.
\end{abstract}

{\bf Keywords:} Bose-Fermi mixtures, Liouville integrability,
Differential Galois groups, Ziglin-Morales-Ramis approach

{\bf 2010 MSC:} 70H05, 70H07,  37J30

\section{Introduction}

In this paper we study the integrability of the system that comes
from the time dependent mean field equations of Bose--Fermi
mixture (BFM) in one dimensional optical lattices. The interest in
BFM arises after the discovery of Bose--Einstein Condensates (BEC)
in 1995 and the desire to understand strongly interacting and
strongly correlated systems, with applications in solid state
physics, nuclear physics, astrophysics, quantum computing and
nanotechnologies. For more detailed physical background of BFM we
refer to \cite{kb04,s05,bpv07,bskk06,KGV} and the literature
therein.

At mean field approximation we consider the following $N_{f} + 1$
coupled nonlinear Schr\"odinger equations
\begin{eqnarray}
\label{1.1} & i\hbar \frac{\partial \Psi^{b}}{\partial t} &+
\frac{1}{2 m_{\rm{B}}}
    \frac{\partial^2 \Psi^{b}}{\partial x^2} - V\Psi^{b}- g_{\rm{BB}}
    |\Psi^{b}|^2\Psi^{b}
    - g_{\rm{BF}}\rho_{f}\Psi^{b}=0, \\
   \label{1.2}
& i\hbar \frac{\partial \Psi^{f}_j}{\partial t} &+
    \frac{1}{2 m_{\rm{F}}}\frac{\partial^2 \Psi^{f}_j}{\partial
    x^2}-V\Psi^{f}_j- g_{\rm{BF}}|\Psi^{b}|^2\Psi^{f}_j=0, \quad j=1,\ldots, N_{f},
    \end{eqnarray}
where the wavefunctions $\Psi_j ^f$ describe each of $N_f$
fermions and $\Psi^b$ is the wavefunction for the bosonic
component, $\rho_{f}=\sum\limits_{i=1}^{N_{f}}|\Psi^{f}_{i}|^2$
and $g_{\rm{BB}}, g_{\rm{BF}}, m_{\rm{F}}, m_{\rm{B}}$ are certain
physical constants. In particular, $g_{\rm{BB}}$ and $
g_{\rm{BF}}$ are related with the s-wave collisions for
boson-boson
 and boson-fermion interactions, respectively. The potential $V$ is usually of the form
 $V = V_0 sn ^2 (\alpha x, \kappa)$, where $ sn (\alpha x, \kappa)$ is the Jacobi elliptic sine function.
 In this paper we take $V_0 = 0$ as in \cite{bpv07}.

We are interested in the stationary solutions to the system
(\ref{1.1}), (\ref{1.2}) of the kind
\begin{eqnarray}
\label{1.3} & \Psi^{b}(x, t) &=
q_{0}(x)\exp\left(-i\frac{\omega_{0}}{\hbar}t+
i\Theta_{0}(x)+i\kappa_{0} \right), \\
\label{1.4} & \Psi^{f}_j(x,t) &=
q_{j}(x)\exp\left(-i\frac{\omega_{j}}{\hbar}t+
i\Theta_{j}(x)+i\kappa _{0,j} \right), \quad j = 1,\ldots, N_{f} ,
\end{eqnarray}
where  $\kappa _{0}$, $\kappa _{0,j} $ are constant phases, $q_0,
q_{j}$ and $\Theta_{0}$, $\Theta _j $ are real-valued functions
related by
\begin{gather}
\Theta_{0}(x)={C}_{0}\int_{0}^{x} \frac{dx'}{q^2_{0}(x')}, \qquad
\Theta_{j}(x)={C}_{j}\int_{0}^{x} \frac{dx'}{q^2_{j}(x')}, \quad
j=1,\ldots, N_{f} \label{1.5}
\end{gather}
${C}_{0},{C}_{j}$,  being constants of integration. After
substituting  (\ref{1.3}), (\ref{1.4}) in equations (\ref{1.1}),
(\ref{1.2}) and separating the real and imaginary part we get
\begin{gather}
\label{1.6} \frac{1}{2 m_{\rm{B}}} q_{0}^{3} q_{0xx} - g_{\rm{BB}}
q_{0}^{6} - g_{\rm{BF}} \left(\sum_{i=1}^{N_{f}}q_{i}^2\right)
q_{0}^{4}+\omega_{0} q_{0}^{4}= \frac{C_{0}^{2} }{2m_{\rm{B}}},
\\
\frac{1}{2 m_{\rm{F}}}q_{j}^{3}q_{jxx} - g_{\rm{BF}}q_{0}^2
q_{j}^{4}  + \omega_{j} q_{j}^{4}= \frac{C_{j}^{2} }{2m_{\rm{F}}},
\quad j = 1, \ldots, N_f .\nonumber
\end{gather}

Kostov et al. \cite{KGV} have found plenty of particular
(quasiperiodic, periodic and soliton) solutions to the system
(\ref{1.6}) and therefore, stationary solutions to the system
(\ref{1.1}), (\ref{1.2}). It is natural to ask whether we can
obtain more, that is, for what set of constants the system
(\ref{1.6}) has enough first integrals to be integrable. Note that
when $g_{\rm{BF}} = 0$ the equations separate, i.e., the system is
solvable.

Before giving our main result let first get rid of the inessential
(for integrability) parameters. In what follows we assume that the
parameters $\omega_0, \omega_j, m_{\rm{F}}, m_{\rm{B}},
g_{\rm{BB}}$ are positive since they have an origin from physics,
and $C_0, C_j, g_{\rm{BF}}$ are arbitrary real parameters. We put
$q_0 = \beta \tilde{q}_0, q_j = \alpha \tilde{q}_j, x = \gamma
\tilde{x}$. Then we choose $\alpha = \sqrt{m_{\rm{F}}}, \beta =
\sqrt{m_{\rm{B}}}, \gamma = 1/(m_{\rm{B}} \sqrt{g_{\rm{BB}}}), \,
g_{\rm{BB}} \neq 0$. Denoting $\tilde{g}_{\rm{BF}} = g_{\rm{BF}}
\alpha^2 \gamma^2 m_{\rm{B}}, \tilde{\omega}_0 = \omega_0 \gamma^2
m_{\rm{B}}, \tilde{\omega}_j = \omega_j \gamma^2 m_{\rm{F}},
\tilde{C}_j ^2 = C_j ^2 \gamma^2 /\alpha^4, \tilde{C}_0 ^2 = C_0
^2 \gamma^2 /\beta^4 $ we reach
\begin{gather}
\label{1.7} \frac{1}{2} \frac{d^2 \tilde{q}_{0}}{d \tilde{x}^2} -
\tilde{q}_{0}^{3} - \tilde{g}_{\rm{BF}} \left( \sum_{i=1}^{N_{f}}
\tilde{q}_{i}^2 \right) \tilde{q}_{0} + \tilde{\omega}_{0}
\tilde{q}_{0} = \frac{\tilde{C}_{0}^{2}}{2 \tilde{q}_0 ^3},
\\
\frac{1}{2} \frac{d^2 \tilde{q}_{j}}{d \tilde{x} ^2} -
\tilde{g}_{\rm{BF}} \tilde{q}_{0}^2 \tilde{q}_{j}  +
\tilde{\omega}_{j} \tilde{q}_{j} = \frac{\tilde{C}_j ^2 }{2
\tilde{q}_j ^3}, \quad j = 1, \ldots, N_f. \nonumber
\end{gather}
To simplify notations we skip the tildas, write $t$ instead of $x$
and denote $p_j = \dot{q}_j, j = 0, \ldots, N_f$, ($. = d/d t$).
Then the system (\ref{1.7}) can be presented as  a Hamiltonian
system with the Hamiltonian

\begin{equation}
\label{1.8} H = \frac{p_0 ^2}{2} + \frac{1}{2} \sum_1 ^{N_f} p_j
^2 + \omega_0 q_0 ^2 + \sum_1 ^{N_f} \omega_j q_j ^2 -g_{\rm{BF}}
q_0 ^2 \sum_1 ^{N_f} q_j ^2 - \frac{q_0 ^4}{2} + \frac{C_0 ^2}{2
q_0 ^2} + \frac{1}{2} \sum_1 ^{N_f} \frac{C_j ^2}{q_j ^2} .
\end{equation}

For the Hamiltonian system with the Hamiltonian (\ref{1.8}) we
consider the cases:

1) $C_0 = 0, C_j \neq 0, \omega_j = \omega^2/2, j = 1, \ldots,
N_f$ ;

2) $C_0 \neq 0, C_j = 0,  j = 1, \ldots, N_f$ ;

3) $C_0 \neq 0, C_1 \neq 0, N_f = 1, $,  $g_{\rm{BF}}$
sufficiently small.

Our  result is the following:
\begin{thm}
\label{th1} For the cases given above, the Hamiltonian system
corresponding to (\ref{1.8}) is non-integrable in Liouville sense
unless  $g_{\rm{BF}} = 0$.
\end{thm}

In other words, the Hamiltonian system under consideration is
integrable only when it is separable.

The proof of the above result is based on the Differential Galois
approach and Ziglin-Morales-Ramis method. This method has been
applied for the studying  the integrability to a number of
Hamiltonian systems, in particular systems with homogeneous
potentials, see \cite{M,MRS1,MR2,PrS}. The classification of all
integrable two degrees of freedom systems with polynomial
potentials of degree 3 is obtained in \cite{MP}. In particular,
the above mentioned approach is used in \cite{AcBl} for obtaining
non-integrability results for some two degrees of freedom
Hamiltonians with rational potentials. Note that the system in
this paper is not of that kind.

For the natural Hamiltonian systems with two degrees of freedom,
similar to (\ref{1.8})
$$
H = \frac{p_1 ^2 + p_2 ^2}{2} + U (q_1, q_2)
$$
there is an integrable generalization of Garnier's system found by
Wojciechowski \cite{Wo85}, namely
$$
U = A q_1 ^2 + B q_2 ^2 + (q_1 ^2 + q_2 ^2)^2 + \frac{C}{q_1 ^2} +
\frac{D}{q_2 ^2} ,
$$
with a rational first integral depending on $A, B, C, D$ (see also
\cite{Pere}). Note that in the system under consideration, the
symmetry is lost, so it is natural to expect integrability only in
the separable case.

The paper is organized as follows.  In the next section we recall
some facts about Differential Galois groups and Morales-Ramis
method which we use. Then, Section 3 is devoted to the proof of
Theorem 1. We finish with some comments.

\section{Differential Galois Theory and Integrability}

Here we summarize some notions and results related to
Ziglin-Morales-Ramis theory.

A differential field $Ê$ is a field with derivation $\partial =
'$, i.e. an additive mapping satisfying Leibnitz rule. A
differential automorphism of $K$ is an automorphism commuting with
the derivation.

Consider a linear system
\begin{equation}
\label{2.1} \dot{x} = A (t) x, \quad x \in \mathbb{C}^n
\end{equation}
with $t$ defined on some Riemann surface. Denote the coefficient
field in (\ref{2.1}) by $K$. Let $x_{i j}$ be the elements of the
fundamental matrix $X (t)$. Let $L (x_{i j})$ be the extension of
$K$ generated by $K$ and $x_{i j}$ -- a differential field. This
extension is called Picard-Vessiot extension.
 Similarly to classical Galois Theory we define the Galois group
$G := Gal_{K} (L) = Gal (L/K)$ to be the group of all differential
automorphisms of $L$ leaving the elements of $K$ fixed. The Galois
group is, in fact, an algebraic group. It has a unique connected
component $G^0$ which contains the identity and which is a normal
subgroup of finite index. The Galois group $G$ can be represented
as an algebraic linear subgroup of $GL (n, \mathbb{C})$ by
$$
\sigma (X (t)) = X (t) R_{\sigma},
$$
where $\sigma \in G$  and $R_{\sigma} \in GL (n, \mathbb{C})$ (see
e.g. \cite{SvP}).

Consider now a Hamiltonian system
\begin{equation}
\label{2.2} \dot{x} = X_{H} (x), \quad t \in \mathbb{C}, \quad x
\in M
\end{equation}
corresponding to an analytic Hamiltonian $H$, defined on the
complex $2 n$-dimensional manifold $M$. Suppose the system
(\ref{2.2}) has a non-equilibrium solution $\Psi (t)$. Denote by
$\Gamma$ its phase curve. We can write the equation in variation
(VE) along this solution
\begin{equation}
\label{2.3} \dot{\mathbf{\xi}} = D X_{H} ( \Psi (t)) \mathbf{\xi},
\quad \mathbf{\xi} \in T_{\Gamma} M.
\end{equation}
Further, using the integral $d H$ we can reduce the variational
equation. Consider the normal bundle of $\Gamma$, $F:= T_{\Gamma}
M / TM$ and let $\pi : T_{\Gamma} M \to F$ be the natural
projection. The equation (\ref{2.3}) induces an equation on $F$
\begin{equation}
\label{2.4} \dot{\eta} = \pi_{*} (D X_{H} ( \Psi (t))(\pi^{-1}
\eta) , \quad \eta \in F.
\end{equation}
which is  called the normal variational equation (NVE).

It is natural to  assume that if the system (\ref{2.2}) is
integrable, then the linear equations (VE) and (NVE) are also
integrable.

 The solutions of (\ref{2.3}) define an extension $L_1$ of the coefficient field
 $K$ of (VE). This naturally defines a differential Galois group
 $G = Gal(L_1/K)$.
  Then, the following  result has established
 \begin{thm}
 \label{th2}
 ({\rm Morales-Ruiz-Ramis \cite{M}}) Suppose that a Hamiltonian system has $n$
  meromorphic first integrals in involution. Then the identity component  $G^0$
  of the Galois group $G = Gal (L_1/K)$ is abelian.
 \end{thm}
Once it is proven, that $G^0$ is not abelian,  the respective
Hamiltonian  system  is non-integrable in the Liouville sense.
Note that the fact that $G^0$ is abelian  doesn't imply
necessarily integrability of the Hamiltonian system. Thus,  one
needs other obstructions to the integrability. A method based on
the higher variational equations has been introduced in \cite{M}
and  the previous Theorem has been extended in \cite{MRS1}. Before
formulating this result let us give an idea of higher variational
equations. For the system (\ref{2.2}) with a particular solution
$\Psi (t)$ we put
\begin{equation}
\label{2.5} x = \Psi (t) + \varepsilon \xi^{(1)} + \varepsilon^2
\xi^{(2)} + \ldots + \varepsilon^k \xi^{(k)} + \ldots,
\end{equation}
where $\varepsilon$ is a formal small parameter. Substituting the
above expression into Eq. (\ref{2.2}) and comparing terms with the
same order in $\varepsilon$ we obtain the following chain of
linear non-homogeneous equations
\begin{equation}
\label{2.6} \dot{\xi}^{(k)} = A (t) \xi^{(k)} + f_k (\xi^{(1)},
\ldots, \xi^{(k-1)}), \quad k = 1, 2, \ldots ,
\end{equation}
where $A (t) = D X_{H} ( \Psi (t))$ and $f_1 \equiv 0$. The
equation (\ref{2.6}) is called k-th variational equation
(${\rm{VE}}_k$). Let $X (t)$ be the fundamental matrix of
(${\rm{VE}}_1$)
$$
\dot{X} = A (t) X .
$$
Then the solutions of $({\rm{VE}}_k), k > 1$ can be found by
\begin{equation}
\label{2.7} \xi^{(k)} = X (t) c (t),
\end{equation}
where $c (t)$ is a solution of
\begin{equation}
\label{2.8} \dot{c} = X^{-1} (t) f_k .
\end{equation}
Although (${\rm{VE}}_k$) are not actually homogeneous equations,
they can be put in that frame,  and therefore, one can define
successive extensions $ K \subset L_1 \subset L_2 \subset \ldots
\subset L_k$, where $L_k$ is the extension obtained by adjoining
the solutions of (${\rm{VE}}_k$). Correspondingly one can define
the Galois groups $Gal (L_1 /K), \ldots, Gal (L_k /K)$. The
following result is proven in \cite{MRS1}.
\begin{thm}
\label{th3} If the Hamiltonian system (\ref{2.2}) is integrable in
Liouville sense then the identity component of every Galois group
$ Gal (L_k /K)$ is abelian.
\end{thm}

Note that we apply Theorem \ref{th3} in the situation when the
identity component of the Galois group $Gal(L_1/K)$ is abelian.
This means that the first variational equation is solvable. Once
we have the solution of $({\rm{VE}}_1)$, then the solutions of
$({\rm{VE}}_k)$ can be found by the method of variations of
constants as explained above. Hence, the Galois groups $Gal(L_k
/K)$ are solvable. One possible way to show that some of them is
not commutative is to find a logarithmic term in the corresponding
solution (see detailed descriptions and explanations in
\cite{M,MRS1,MR2}).

%%%%%%%%%%%%%%%%%%%%%%%%%%%%%%%

Now we recall a perturbational technique which is still related to
the Differential Galois approach. Let $M_0$ be a two-dimensional
complex analytic symplectic manifold, $H_0 (q, p)$ be a
holomorphic Hamiltonian and $X_{H_0}$ be the corresponding
Hamiltonian vector field. Assume that the system
\begin{equation}
\label{2.101} \dot{q} = H_{0, p} , \quad \dot{p} = -H_{0, q}
\end{equation}
has a hyperbolic equilibrium $(q_0, p_0)$. Then the system
(\ref{2.101}) has a separatrix
\begin{equation}
\label{2.102} \Gamma_0 : (q_0 (t), p_0 (t)), \lim_{t \to \infty}
q_0 (t) = q_0 \, , \lim_{t \to \infty} p_0 (t) = p_0 .
\end{equation}
The functions $q_0 (t) , p_0 (t)$ are meromorphic in $t \in
\mathbb{C}$. Let
\begin{equation}
\label{2.103} H (q, p, t, \varepsilon) = H_0 (q, p) + \varepsilon
H_1 (q, p, t) + \ldots
\end{equation}
be a meromorphic small (complex) perturbation of $H_0$ satisfying
$H_1 (q, p, t + \omega) = H_1 (q, p, t)$ with  a period $\omega
\in \mathbb{C}$. This function $H$ is defined over $M = M_0 \times
F_{\omega}, \, F_{\omega} = \mathbb{C}/\omega \mathbb{Z}$. We can
write the Hamiltonian system defined by $H (q, p, \varphi)$ over
$M$ as
\begin{equation}
\label{2.104} \dot{q} = H_{p} , \quad \dot{p} = -H_{q}, \quad
\dot{\varphi} = 1, \quad (q, p, \varphi) \in M.
\end{equation}
When $\varepsilon = 0$ the system (\ref{2.104}) reduces to
\begin{equation}
\label{2.105} \dot{q} = H_{0, p} , \quad \dot{p} = -H_{0, q},
\quad \dot{\varphi} = 1, \quad (q, p, \varphi) \in M.
\end{equation}
The unperturbed system (\ref{2.105}) has a hyperbolic
$\omega$-periodic orbit $\Pi_0 := (q_0, p_0 , \varphi = t (\mod
\omega))$. It is well known that for small $|\,\varepsilon|$ the
perturbed system (\ref{2.104}) has also an $\omega$-periodic orbit
$\Pi_{\varepsilon} := (q (t, \varepsilon) , p (t, \varepsilon) ,
\varphi =  t - t_0 (\mod \omega))$, such that $(q (t, 0) , p (t,
0)) = (q_0 , p_0 )$.

We define  the (stable) complex separatrix $\Lambda_{\varepsilon}
^+ $ of the system (\ref{2.104}) as the set of integral curves of
(\ref{2.104}) asymptotic to $\Pi_{\varepsilon}$ as $t \to \infty$.
For fixed $\varepsilon$, it is a two-dimensional complex surface.
This separatrix can have transverse self-intersections.

{\bf Remark 1}. Recall that in the real case the separatrices can
not have transverse self-intersections. Such intersections can
occur between stable and unstable separatrices. For real
Hamiltonian systems, the existence of such transverse orbits is
considered as a source of chaotic behavior and is an obstruction
to existence of an analytic first integral.

Ziglin \cite{Z3} proved that for complex Hamiltonian systems, the
existence of transverse self-intersections for separatrices is
also an obstruction to the integrability.

The unperturbed separatrix is given by $\Lambda_{0} ^+ = \Gamma_0
\times F_{\omega}$. It is foliated by the one-parameter family of
integral curves
\begin{equation}
\label{2.106} \Gamma_{t_0} : (q_0 (t), p_0 (t), t - t_0) ,
\end{equation}
$t_0 \in F_{\omega}$ being the parameter. Let $\gamma : [0, 1] \to
\mathbb{C}$ be a closed path in the complex plane with $\gamma (0)
= \gamma (1) \in \mathbb{R} \subset \mathbb{C}$.  The following
function on $F_{\omega}$
\begin{equation}
\label{2.107} d (t_0) := \int_{\gamma} \{H_0, H_1 \} (q_0 (t), p_0
(t), t - t_0) d t
\end{equation}
is usually called Poincar\'e-Arnold-Melnikov integral. Here $\{ ,
\}$ is the Poisson bracket. Then the following result is valid:
\begin{thm}
\label{th4} (Ziglin) If the function $d (t_0)$ has a simple zero,
then for sufficiently small $|\,\varepsilon| \neq 0$, the
separatrix $\Lambda_{\varepsilon} ^+ $ has a transversal
self-intersection and the system (\ref{2.104}) has no additional
holomorphic first integral.
\end{thm}

It appears that there is a relation between Theorem 2 and Theorem
4. Morales-Ruiz \cite{M3} proved that, under certain assumptions,
the Ziglin's condition about the Poincar\'e-Arnold-Melnikov
integral can be interpreted by the fact that the Galois group of
the perturbed variational equation along the integral curve
$\Gamma_0$ is non-abelian. In other words, if
Poincar\'e-Arnold-Melnikov integral $d (t_0)$ is not identically
zero, the Galois group of the perturbed variational equation is
not abelian and the system is not integrable by means of
meromorphic first integrals.

\section{Proof of  Theorem 1}

In what follows we assume that $t, q_0 (t), q_j (t)$ are complex
quantities, but we keep the parameters real. The proof goes in the
following lines. For the first two cases we find particular
solutions. Then we study the variational equation (VE) along these
solutions. The first case is the simplest, that is why we start
with it. The variational equation (VE) is reduced to a particular
case of double confluent Heun equation, which Galois group is more
or less known.

The second case needs more steps. The identity component of the
Galois group of (VE) is not commutative except for some discrete
values of $g_{\rm{BF}}$. By studying higher variational equations
we find a logarithmic term in solutions of (VE$_2$) and (VE$_3$)
when $g_{\rm{BF}} \neq 0$, which implies non commutativity of the
identity component of $Gal (L_2 /K)$ ($Gal (L_3 /K)$) and hence,
non-integrability of our Hamiltonian system.

For the third case we use a perturbational technique which is
still related to the Differential Galois approach. We study the
Poincar\'e-Arnold-Melnikov integral in order to show that a
complex separatrix self-intersects.

\subsection{The case $C_0 = 0, C_j \neq 0$.}

In this case the Hamiltonian (\ref{1.8}) becomes
\begin{equation}
\label{4.1} H = \frac{p_0 ^2}{2} + \frac{1}{2} \sum_1 ^{N_f} p_j
^2 + \omega_0 q_0 ^2 + \sum_1 ^{N_f} \omega_j q_j ^2 -g_{\rm{BF}}
q_0 ^2 \sum_1 ^{N_f} q_j ^2 - \frac{q_0 ^4}{2} + \frac{1}{2}
\sum_1 ^{N_f} \frac{C_j ^2}{q_j ^2} .
\end{equation}
The equations corresponding to the Hamiltonian (\ref{4.1}) are
\begin{eqnarray}
\label{4.109}
\dot{q_0} = p_0, &  & \dot{p_0}=-2\omega_0 q_0 + 2 q_0^3 + 2 g_{\rm{BF}}  q_0\sum_1^{N_f}{q_j}^2, \nonumber \\
\dot{q_j} = p_j, &  & \dot{p_j}=-2\omega_j q_j + 2 g_{\rm{BF}}
q_0^2 q_j+\frac{C_j^2}{q_j^3}, \quad j = 1, \dots, N_f.
\end{eqnarray}

\begin{prop}
\label{prop1} The  system (\ref{4.109}) has a particular solution
of the form
\begin{eqnarray}
\label{4.1010}
q_0 = p_0 = 0, &  & \nonumber \\
{q_j}^2 = \frac{C_j}{\sqrt{2\omega_j}}\sinh(2i\sqrt{2\omega_j} t),
&  & p_j = \dot{ q_j}, \quad  j = 1, \dots, N_f.
\end{eqnarray}
\end{prop}
{\bf Proof.} We put $q_0 = p_0 = 0$ in (\ref{4.109}). The general
solution of the system with respect to $(q_j, p_j), j= 1, \ldots,
N_f$ is
\begin{equation}
\label{4.10101} {q_j}^2 = \frac{h_j}{2\omega_j} + \sqrt{\frac{C_j
^2}{2 \omega_j} - \frac{h_j ^2}{4 \omega_j ^2} } \sinh 2 i \sqrt{2
\omega_j} (t-t_0), \quad  p_j = \dot{ q_j}, \quad  j = 1, \dots,
N_f,
\end{equation}
here $h_j$ are arbitrary constants. Then we set $h_j = 0$ and $t_0
= 0$ to obtain our particular solution.

$\hfill \square$

Denote the variations by $\xi_0 = d q_0$ and $\eta_0 = d p_0$. It
is easy to be seen that the (NVE) are written in variables
$\xi_0$, $\eta_0$, namely
\begin{eqnarray}
\label{4.2} \dot{\xi_0} = \eta_0, & & \dot{\eta}_0
=\left[-2\omega_0 + 2 g_{\rm{BF}} \sum_1^{N_f}{{q_j}^2}\right]
\xi_0.
\end{eqnarray}
We rewrite (\ref{4.2}) as a second order equation
\begin{equation}
\label{4.201} \ddot{\xi}_0 + \left[2\omega_0 - 2 g_{\rm{BF}}
\sum_1^{N_f}{\frac{C_j}{\sqrt{2\omega_j}}\sinh(2i\sqrt{2\omega_j}
t)}\right] \xi_0 = 0.
\end{equation}
The study of the identity component of the Galois group of
(\ref{4.201}) is a difficult task. That is why we assume that all
$\omega_j$ are equal. We put $\omega_j = \frac{\omega^2}{2}$, $j =
1, \dots, {N_f}$. Then we get a variant of Mathieu equation
\begin{equation}
\label{4.2011} \ddot{\xi}_0 + \left[A_1 + B_1 \sinh(2i \omega
t)\right] \xi_0 = 0,
\end{equation}
where
\begin{equation}
\label{4.2012} A_1 = 2 \omega_0,\qquad B_1 = -\frac{2}{\omega}
g_{\rm{BF}} \sum_1^{N_f}{C_j}.
\end{equation}
Since $C_j$ are constants of integration, we can always assume
that $\sum C_j \neq 0$.

Next, by changing the independent variable $x = e^{2i\omega t}$ we
get an algebraic version of (\ref{4.2011})
\begin{equation}
\label{4.2013} \xi_0 '' + \frac{1}{x}\xi_0 ' + \left[\frac{B}{x} +
\frac{A}{x^2} - \frac{B}{x^3}\right] \xi_0 = 0,
\end{equation}
where $^{'} =\frac{d}{dx}$, $A = -\frac{A_1}{4 \omega^2}$, $B =
-\frac{B_1}{8\omega^2}$. It is obvious that when $B = 0$ this
equation becomes an Euler equation which is solvable. Further, we
reduce (\ref{4.2013}) to the standard form by putting $y =
\sqrt{x} \xi_0$,
\begin{equation}
\label{4.2014} y'' = r (x) y, \qquad r (x) = - \frac{B}{x} -
\frac{A + \frac{1}{4}}{x^2} + \frac{B}{x^3}.
\end{equation}
The equation (\ref{4.2014}) is a particular case of double
confluent Heun equation. For this equation the points $0$ and
$\infty$ are irregular singular ones and one natural way to study
the Galois group is the Kovacic algorithm. This is done by A.
Duval and M. Loday-Richaud in \cite{ADMLR} p.237. We just apply
their result which simply says that if $B \ne 0$ the Galois group
of (\ref{4.2014}) is ${\rm SL} (2, \mathbb{C})$. In our case
$$B = \frac{g_{\rm{BF}}}{4\omega^3} \sum_1^{N_f}C_j,$$
which means that under the assumption $\sum_1^{N_f} C_j \neq 0$
$$B = 0 \, \Leftrightarrow \, g_{\rm{BF}} = 0, $$
that is, the identity component of the Galois group is
noncommutative if $g_{\rm{BF}} \neq 0$. Therefore, by Theorem 1
the Hamiltonian system (\ref{4.1}) is non-integrable unless
$g_{\rm{BF}} = 0$. This finishes the proof of this part  of
Theorem 1.

 {\bf Remark 2.} Let us note that in \cite{AcBl,Ac2,AcMRJW} a
systematic procedure is presented, called Hamiltonian
Algebrization, which transforms second order linear differential
equations with non-rational coefficients into differential
equations with rational coefficients. As an example, the Mathieu
equation is considered, see for instance, section 2.1 in
\cite{AcBl}. The conclusion is the same: the Mathieu equation is
not integrable for $B \neq 0$.

\subsection{The case $C_0 \neq 0, C_j = 0$.}

Let us find a particular solution first.
\begin{prop}
\label{prop2} The Hamiltonian system generated by the Hamiltonian
(\ref{1.8}) with $C_j = 0$ has a particular solution in the form
\begin{equation}
\label{3.1} \bar{q}_0^2 (t) = \frac{2}{3} \omega_0 + \wp (t; g_2,
g_3), \quad \bar{p}_0 (t) = \dot{\bar{q}}_0 (t), \quad q_j = p_j =
0, \quad j = 1,\dots, N_f,
\end{equation}
where $\wp (t; g_2, g_3)$ is the Weierstrass elliptic function
satisfying
\begin{equation}
\label{3.2} \Gamma : \dot{v}^2 = 4 v^3 - g_2 v - g_3
\end{equation}
with $g_2 = \frac{16}{3}\omega_0^2 - 4 h$, $g_3 = 4 C_0^2 -
\frac{8}{3}\omega_0 h + \frac{64}{27} \omega_0^3$ and $h$ is level
of the Hamiltonian (\ref{1.8}), chosen so that $\Delta = g_2^3 -
27 g_3^2 \neq 0$.
\end{prop}
\noindent {\bf Proof}. We put $q_j = p_j = 0, \, j = 1, \ldots,
N_f$ (recall $C_j = 0$) in (\ref{1.8}) to obtain
\begin{equation}
\label{3.21} H = \frac{p_0 ^2}{2} + \omega_0 q_0 ^2 - \frac{q_0
^4}{2} + \frac{{C_0 ^2}}{2 q_0 ^2} = \frac{h}{2} .
\end{equation}
We rewrite this  expression in the form
\begin{equation}
\label{3.22} \dot{q}_0 ^2  = - 2 \omega_0 q_0 ^2 + q_0 ^4 -
\frac{{C_0 ^2}}{ q_0 ^2} + h.
\end{equation}
Then denoting $u = q_0 ^2 $ and also $u = v + \frac{2}{3}
\omega_0$ we obtain the general solution of (\ref{3.22})
\begin{equation}
\label{3.222} \bar{q}_0^2 (t) = \frac{2}{3} \omega_0 + \wp (t-t_0;
g_2, g_3), \quad \bar{p}_0 (t) = \dot{\bar{q}}_0 (t) .
\end{equation}
We set $t_0 = 0$ to get the desired result.

$\hfill \square$

%%%%%{\bf Remark 1.} Note that $-\frac{2}{3}\omega_0$ is not a root of the equation
%%%%% $4v^3 - g_2 v - g_3 = 0$ since $C_0 \neq 0$.

Next we write the variational equations (VE) along the particular
solution (\ref{3.1}). Denote $\xi_0 = dq_0$, $\eta_0 = dp_0$,
$\xi_j = dq_j$, $\eta_j = dp_j$. Then the (VE) can be written as
\begin{eqnarray}
  \label{3.3}
   & \dot{\xi_0}  = \eta_0, \qquad &
    \dot{\eta_0} = \left(-2\omega_0 + 6 \bar{q}_0^2(t)-\frac{3{C}_0^2}{\bar{q}_0 ^4(t)}\right)\xi_0 , \\
  \label{3.4}
   & \dot{\xi}_j   = \eta_j, \qquad &
  \dot{\eta}_j = \left(-2\omega_j + 2 g_{\rm{BF}}\bar{q}_0 ^2 (t) \right) \xi_j , \quad j = 1, \dots, N_f .
\end{eqnarray}
The equation (\ref{3.3}) forms the tangent part of (VE) and the
equations (\ref{3.4}) form the normal part of (VE), actually
(NVE). It is seen from (\ref{3.4}) that (NVE) splits into a system
of $N_f$ independent equations (NVE$_j$), $j = 1, \ldots, N_f$.
Hence, (NVE) is integrable if, and only if, each of (NVE$_j$) is
integrable. In other words, the identity component of the Galois
group of (NVE) is solvable (commutative) if, and only if, each of
identity components of the Galois groups of the (NVE$_j$) is
solvable (commutative). Therefore, it is enough to study one of
them. Let us write (NVE$_j$) for certain particular $j$ as a
second order equation
\begin{equation}
\label{3.5} \ddot{\xi}_{j} + \left(2 \omega_j -2
g_{\rm{BF}}\bar{q}_0 ^2 (t) \right) \xi_j = 0.
\end{equation}
Taking into account the particular solution (\ref{3.1}) the Eq.
(\ref{3.5}) is a Lam\'e equation
\begin{equation}
\label{3.6} \ddot{\xi}_{j} + \left(2\omega_j -\frac{4}{3}
g_{\rm{BF}}\omega_0-2g_{\rm{BF}}\wp(t)\right) \xi_j = 0.
\end{equation}
It can be proven that if $g_{\rm{BF}} \neq \frac{n(n+1)}{2}$, $n
\in \mathbb{Z}$ the monodromy group of (\ref{3.6}) is not Abelian
(see e.g. \cite{M}). Since the equation (\ref{3.6}) is a Fuchsian
one, the monodromy group generates the differential Galois group,
and hence the Galois group is not abelian. Then due to  Theorem
\ref{th2} the Hamiltonian system with the Hamiltonian (\ref{1.8})
is non integrable in Liouville sense.

 Further, we study the tangential part of the (VE) - Eq. (\ref{3.3}). The theory gives that its Galois group is solvable.
In fact, we have
\begin{prop}
\label{prop3} The Galois group of (\ref{3.3}) is abelian.
\end{prop}
\noindent {\bf Proof}. It is well known that the system
(\ref{3.3}) has a particular solution $(\xi_{0,1},
\dot{\xi}_{0,1}) = (\bar{p_0} (t),  \dot{\bar{p_0}} (t))$. The
other solution is obtained via D'Alembert's formula
$$
\xi_{0,2} = \xi_{0,1} \int_0 ^t \frac{d \tau}{(\xi_{0,1})^2} .
$$

Denote  the coefficient field of (\ref{3.3}) by $K = \mathbb{C}
\left(\wp (t),\wp' (t) \right)$. This field is  isomorphic to the
field of meromorphic functions $\mathcal{M} (\Gamma)$ on
$\Gamma$.

It can be seen from the obtained solutions  that one part of them
lie in a quadratic extension of the field $K$ and the another part
is obtained with single quadrature of the elements of this
extension. Therefore the Galois group of (\ref{3.3}) acts in the
following way: $\sigma \in Gal(L/ K)$, $\sigma(\xi_{0,1}) =
\xi_{0,1}$ and $\sigma (\xi_{0,2}) = \xi_{0,2} + \nu_{0}
\xi_{0,1}$, $\alpha_0 \in \mathbb{C}$. Let $\Xi (t)$ is the
fundamental matrix of (\ref{3.3})
$$
\Xi = \left( \begin{array}{cc}
\xi_{0,2} & \, \xi_{0,1} \\
\dot{\xi}_{0,2} & \, \dot{\xi}_{0,1}
\end{array}\right).
$$
Then $\sigma \in Gal(L/ K)$ can be represented by the matrix
$R_{\nu_{0}}$,  $\sigma \Xi (t) = \Xi (t) R_{\nu_{0}}$, where
 $R_{\nu_{0}}=
 \left(\begin{array}{cc}
 1 & 0 \\
 \nu_{0} & 1
 \end{array}\right)$.
 It is clear that the group
 $\left\{
 \left(
 \begin{array}{cc}
 1 & 0\\
 \nu_{0} & 1
 \end{array}\right)
 \right\}
  $
  is abelian.

$\hfill \square$

 So far we have shown that if $g_{\rm{BF}} \neq \frac{n(n+1)}{2}, n \in \mathbb{Z}$ the identity component of the Galois group of
 (NVE) is not abelian and hence the Hamiltonian system under consideration is non-integrable.

Now, let us consider the case when
\begin{equation}
\label{3.7} g_{\rm{BF}} =  \frac{n(n+1)}{2},\qquad n \in
\mathbb{Z}.
\end{equation}
Then every equation (\ref{3.6}) is a Lam\'e equation in
Weierstrass form
\begin{equation}
\label{3.71} \ddot{\xi}_j - \left[ n (n+1) \wp (t) + B_j \right]
\xi_j = 0,
\end{equation}
where $B_j = \frac{2}{3} \omega_0 n (n+1) - 2 \omega_j$. The cases
for which the Lam\'e equation (\ref{3.71}) is solvable are well
known:

(i) The Lam\'e and Hermite solutions. In this case $n \in
\mathbb{Z}$ and $g_2, g_3, B$ are arbitrary parameters;

(ii) The Brioschi-Halphen-Crowford solutions. Here $m:= n + 1/2
\in \mathbb{N}$ and the parameters $g_2, g_3, B$ must satisfy an
algebraic equation.

(iii) The Baldassarri solutions. Now $n + 1/2 \in \frac{1}{3}
\mathbb{Z} \cup \frac{1}{4} \mathbb{Z} \cup \frac{1}{5} \mathbb{Z}
\setminus \mathbb{Z}$ with additional algebraic relations between
the other parameters.

Note that in the case (i) the identity component of the Galois
group $G^0$ is of the form $\begin{pmatrix}
1 & 0 \\
\nu_j & 1
\end{pmatrix}$
 and in the cases (ii) and (iii) $G^0 = id$ ($G$ is finite).
And these are the all cases when the Lam\'e equation is
integrable.

Therefore, together with the result of Proposition \ref{prop2} we
have that the identity component of Galois group of the (VE) is
 represented by the block-diagonal matrices of the kind
  $$
  \begin{pmatrix}

 \begin{matrix}1 & 0\\ \nu_{0} & 1 \end{matrix} & \begin{matrix}0 & 0 \\ 0 & 0 \end{matrix} & \dots &  \begin{matrix}0 & 0 \\ 0 & 0 \end{matrix} \\

 \ldots & \ldots & \ldots & \ldots \\
 \begin{matrix}0 & 0\\ 0 & 0 \end{matrix} & \begin{matrix}1 & 0 \\ \nu_{j} & 1 \end{matrix} & \dots &  \begin{matrix}0 & 0 \\ 0 & 0 \end{matrix} \\
 \ldots & \ldots & \ldots & \ldots \\
 \begin{matrix}0 & 0\\ 0 & 0 \end{matrix} & \begin{matrix}0 & 0\\ 0 & 0 \end{matrix} & \dots & \begin{matrix}1 & 0\\ \nu_{N_f} & 1 \end{matrix}

  \end{pmatrix}
  $$
  and it is clearly commutative.

%%%%%%%%%%%%%%%%%%%%%%

The integrability of Hamiltonian systems with two degrees of
freedom which (NVE) are Lam\'e equations is studied in
\cite{MSim,M}. We summarize the facts and the result (Theorem 5),
that gives necessary conditions for integrability in the Appendix.
Since in our case the (NVE) splits into a system of $N_f$
equations, the result for two degrees of freedom can be applied.

The potential $\varphi (q_0)$ is obtained from (\ref{3.22}).
Denote $\alpha (t, h) : = n (n+1) \wp (t) + B_j$. We calculate the
coefficients of the polynomial $P (\alpha, h)$ (compare with (A.4)
and (\ref{coeff}))
$$
P (\alpha, h) = (a_1 + h a_2) \alpha^3 + (b_1 + h b_2) \alpha^2 +
(c_1 + h c_2) \alpha + (d_1 + h d_2) .
$$
In our case these coefficients are:
\begin{eqnarray}
& a_1 = \frac{4}{n(n+1)}, \quad a_2 = 0, \quad b_1 = -\frac{12 B_j}{n (n+1)}, \quad b_2 = 0 \nonumber \\
& c_1 = \frac{12 B_j ^2}{n (n+1)} - \frac{16}{3} \omega_0 ^2 n (n+1), \quad c_2 = 4 n (n+1), \\
& d_1 = \frac{16}{3} b_j \omega_0 ^2 n (n+1) - \frac{4}{n (n+1)}
B_j ^3 - n^2 (n+1)^2 \left( 4 C_0 ^2 + \frac{64}{27} \omega_0 ^3
\right) , \quad d_2 = 8 n (n+1) \omega_j . \nonumber
\end{eqnarray}
Now we are ready to apply Theorem 5 (see the Appendix).

The condition 3 is not fulfilled: $c_2 \neq 0$ and $c_2 b_1 - 3
a_1 d_2 = - 32 \omega_0 n (n+1)$,  which is nonzero by the
assumption that $\omega_0, \omega_j$ are positive numbers, made in
the very beginning. In particular, there are no Baldassarri
solutions.

We proceed with the cases of the condition 2. In the case 2.1, $m
= 1$, $b_1 = 0 $ is equivalent to
\begin{equation}
\label{3.711} B_j = 0, \quad \mbox{or}, \quad \omega_j = \omega_0
/ 4, \quad j = 1, \ldots, N_f.
\end{equation}
If for some $j$ $ B_j \neq 0$ then the system is not integrable
for this $m$. We will consider the case when all $B_j =0$ in what
follows.

The case 2.2 $m = 2$ does not occur here since $c_2 \neq 0$.

In the case 2.3, $m=3$ the necessary conditions
$$
16 a_1 d_2 + 11 b_1 c_2 = 0, \quad  1024 a_1 ^2 d_1 + 704 a_1 b_1
c_1 + 45 b_1 ^3 = 0
$$
yield correspondingly
\begin{equation}
\label{3.712} B_j = \frac{32}{33} \omega_j, \quad (55 \omega_0 =
28 \omega_j), \quad 7^3 C_0 ^2 = 72 \omega_0 ^3 .
\end{equation}
If any of the above conditions is violated, then the system is
non-integrable for this $m$. We will consider the case when
relations (\ref{3.712}) are valid for all $j$ in what follows.

Finally, the case 2.m, $m > 3$ does not occur here since $c_2 \neq
0$ and $d_2 \neq 0$.

\vspace{2ex}

In order to resolve the condition 1, the case 2.1 with
(\ref{3.711}) and the case 2.3 with (\ref{3.712}) of the condition
2 in the Theorem 5 we need to study the Galois groups of higher
variational equations and to apply Theorem 3. To compute higher
variations we put
\begin{eqnarray}
\label{3.100}
q_0 & = & \bar{q}_0  +  \varepsilon \xi_0 ^{(1)} + \varepsilon^2 \xi_0 ^{(2)} + \varepsilon^3 \xi_0 ^{(3)} + \ldots, \nonumber\\
p_0 & = & \bar{p}_0 + \varepsilon \eta_0 ^{(1)} + \varepsilon^2 \eta_0 ^{(2)} + \varepsilon^3 \eta_0 ^{(3)} + \ldots, \\
q_j & = & 0      \, \,   + \varepsilon \xi_j ^{(1)} + \varepsilon^2 \xi_j ^{(2)} + \varepsilon^3 \xi_j ^{(3)} + \ldots, \nonumber \\
p_j & = & 0      \, \,  + \varepsilon \eta_j ^{(1)} +
\varepsilon^2 \eta_j ^{(2)} + \varepsilon^3 \eta_j ^{(3)} +
\ldots, \quad j = 1, \ldots, N_f . \nonumber
\end{eqnarray}
and substitute these expressions into the original Hamiltonian
system. Comparing the terms with the same order  in $\varepsilon$,
we get consecutively  the variational equations up to order 3.

The first variational equation is
\begin{eqnarray}
\label{3.101} & \dot{\xi}_0 ^{(1)} = \eta_0 ^{(1)}, \quad &
\dot{\eta}_0 ^{(1)} = \left( - 2 \omega_0 + 6 \bar{q}_0 ^2 - \frac{3 C_0 ^2}{\bar{q}_0 ^4} \right) \xi_0 ^{(1)}, \\
\label{3.102} & \dot{\xi}_j ^{(1)} = \eta_j ^{(1)}, \quad &
\dot{\eta}_j ^{(1)} = ( - 2 \omega_j + 2 g_{\rm{BF}} \bar{q}_0 ^2)
\xi_j ^{(1)}, \quad j = 1, \ldots, N_f ,
\end{eqnarray}
but of course we know it (see (\ref{3.3}, (\ref{3.4})). For the
second variational equation we have
\begin{eqnarray}
\label{3.103} & \dot{\xi}_0 ^{(2)} = \eta_0 ^{(2)}, \quad &
\dot{\eta}_0 ^{(2)} = \left( - 2 \omega_0 + 6 \bar{q}_0 ^2 - \frac{3 C_0 ^2}{\bar{q}_0 ^4} \right) \xi_0 ^{(2)} + K_0 ^{(2)}, \\
\label{3.104} & \dot{\xi}_j ^{(2)} = \eta_j ^{(2)}, \quad &
\dot{\eta}_j ^{(2)} = ( - 2 \omega_j + 2 g_{\rm{BF}} \bar{q}_0 ^2)
\xi_j ^{(2)} + K_j ^{(2)}, \quad j = 1, \ldots, N_f .
\end{eqnarray}
The third variational equation is
\begin{eqnarray}
\label{3.105} & \dot{\xi}_0 ^{(3)} = \eta_0 ^{(3)}, \quad &
\dot{\eta}_0 ^{(3)} = \left( - 2 \omega_0 + 6 \bar{q}_0 ^2 - \frac{3 C_0 ^2}{\bar{q}_0 ^4} \right) \xi_0 ^{(3)} +  K_0 ^{(3)}, \\
\label{3.106} & \dot{\xi}_j ^{(3)} = \eta_j ^{(3)}, \quad &
\dot{\eta}_j ^{(3)} = ( - 2 \omega_j + 2 g_{\rm{BF}} \bar{q}_0 ^2)
\xi_j ^{(3)} + K_j ^{(3)}, \quad j = 1, \ldots, N_f .
\end{eqnarray}
Here
\begin{eqnarray}
\label{3.107}
K_0 ^{(2)} & = & 2 g_{\rm{BF}} \bar{q}_0 \sum (\xi_j ^{(1)})^2 + 6 \bar{q}_0 (\xi_0 ^{(1)})^2 + 6 C_0 ^2 \frac{ (\xi_0 ^{(1)})^2}{\bar{q}_0 ^5} , \nonumber \\
K_j ^{(2)} & = & 4 g_{\rm{BF}} \bar{q}_0 \xi_0 ^{(1)} \xi_j ^{(1)}, \quad j = 1, \ldots, N_f  ,\nonumber \\
K_0 ^{(3)} & = & 2 g_{\rm{BF}} \left[2 \bar{q}_0 \sum \xi_j ^{(1)}
\xi_j ^{(2)}  + \xi_0 ^{(1)} \sum (\xi_j ^{(1)})^2 \right] +
2  (\xi_0 ^{(1)})^3  + 12 \bar{q}_0 \xi_0 ^{(1)} \xi_0 ^{(2)} \\
& - & \frac{C_0 ^2}{\bar{q}_0 ^6} \left[10 (\xi_0 ^{(1)})^3 - 12 \bar{q}_0 \xi_0 ^{(1)} \xi_0 ^{(2)}  \right] ,    \nonumber                                                          \\
K_j ^{(3)} & = & 2 g_{\rm{BF}} \left[ (\xi_0 ^{(1)})^2 \xi_j
^{(1)} + 2 \bar{q}_0 \left( \xi_0 ^{(1)} \xi_j ^{(2)} + \xi_0
^{(2)} \xi_j ^{(1)} \right)  \right], \quad j = 1, \ldots, N_f.
\nonumber
\end{eqnarray}
Then, in our notation from Section 2, we have
\begin{eqnarray}
\label{3.108}
f_2 & = & \left[0, K_0 ^{(2)}, 0, K_1 ^{(2)}, \ldots, 0, K_{N_f} ^{(2)}  \right]^T, \nonumber \\
f_3 & = & \left[0, K_0 ^{(3)}, 0, K_1 ^{(3)}, \ldots, 0, K_{N_f}
^{(3)}  \right]^T .
\end{eqnarray}

First, we have to solve (${\rm{VE}}_1$). Let $\xi_{0,1} ^{(1)},
\xi_{0,2} ^{(1)}$ be two linearly independent solutions of
(\ref{3.101}) with Wronskian equal to unity, i.e., $ \xi_{0,1}
^{(1)} \dot{\xi}_{0,2} ^{(1)} - \dot{\xi}_{0,1} ^{(1)} \xi_{0,2}
^{(1)} =1$. Similarly, $\xi_{j,1} ^{(1)}, \xi_{j,2} ^{(1)}$ are
linearly independent solutions of (\ref{3.102}) with Wronskian
equal to unity. Then the fundamental matrix $X (t)$ of
(\ref{3.101}), (\ref{3.102}) and its inverse have the
block-diagonal form

%%%%%%%%%%

\begin{equation}
\label{3.1081} X (t) =
\begin{pmatrix}

\begin{matrix} \xi_{0,1} ^{(1)} & \xi_{0,2} ^{(1)}\\ \dot{\xi}_{0,1} ^{(1)} & \dot{\xi}_{0,2} ^{(1)}  \end{matrix} & \begin{matrix}0 & 0 \\ 0 & 0 \end{matrix} & \dots &  \begin{matrix}0 & 0 \\ 0 & 0 \end{matrix} \\

 \ldots & \ldots & \ldots & \ldots \\
 \begin{matrix}0 & 0\\ 0 & 0 \end{matrix} & \begin{matrix} \xi_{j,1} ^{(1)} & \xi_{j,2} ^{(1)} \\ \dot{\xi}_{j,1} ^{(1)} & \dot{\xi}_{j,2} ^{(1)} \end{matrix} & \dots &  \begin{matrix}0 & 0 \\ 0 & 0 \end{matrix} \\
 \ldots & \ldots & \ldots & \ldots \\
 \begin{matrix}0 & 0\\ 0 & 0 \end{matrix} & \begin{matrix}0 & 0\\ 0 & 0 \end{matrix} & \dots & \begin{matrix} \xi_{N_f,1} ^{(1)} & \xi_{N_f,2} ^{(1)}\\ \dot{\xi}_{N_f,1} ^{(1)} & \dot{\xi}_{N_f,2} ^{(1)}  \end{matrix}

  \end{pmatrix},
  \end{equation}
  \begin{equation}
  \label{3.1082}
X^{-1} (t) =
\begin{pmatrix}

 \begin{matrix}\dot{\xi}_{0,2} ^{(1)} & -\xi_{0,2} ^{(1)}\\ -\dot{\xi}_{0,1} ^{(1)} & \xi_{0,1} ^{(1)} \end{matrix} & \begin{matrix}0 & 0 \\ 0 & 0 \end{matrix} & \dots &  \begin{matrix}0 & 0 \\ 0 & 0 \end{matrix} \\

 \ldots & \ldots & \ldots & \ldots \\
 \begin{matrix}0 & 0\\ 0 & 0 \end{matrix} & \begin{matrix} \dot{\xi}_{j,2} ^{(1)} & -\xi_{j,2} ^{(1)}\\ -\dot{\xi}_{j,1} ^{(1)} & \xi_{j,1} ^{(1)}  \end{matrix} & \dots &  \begin{matrix}0 & 0 \\ 0 & 0 \end{matrix} \\
 \ldots & \ldots & \ldots & \ldots \\
 \begin{matrix}0 & 0\\ 0 & 0 \end{matrix} & \begin{matrix}0 & 0\\ 0 & 0 \end{matrix} & \dots & \begin{matrix}\dot{\xi}_{N_f,2} ^{(1)} & -\xi_{N_f,2} ^{(1)} \\ -\dot{\xi}_{N_f,1} ^{(1)} & \xi_{N_f,1} ^{(1)} \end{matrix}

  \end{pmatrix}
\end{equation}

%%%%%%%%%%%%%

The first variational equations (${\rm{VE}}_1$) (\ref{3.101}),
(\ref{3.102})  have a  singular point at $t=0$ (the pole of $\wp
(t)$). We calculate the expansion of the solutions of variational
equations around  the point $t = 0$. Note that
\begin{equation}
\label{3.a1} \bar{q}_0 (t) = \frac{1}{t} + \frac{\omega_0}{3} t +
\left( \frac{g_2}{40} - \frac{\omega_0 ^2}{18} \right) t^3 +
\ldots .
\end{equation}
Here and further dots denote the higher order terms with respect
to $t$. In a neighborhood of $t = 0$ we have the following
expansions for the solutions of the tangential part of
(${\rm{VE}}_1$) Eq. (\ref{3.101})
\begin{equation}
\label{3.109} \xi_{0, 1} ^{(1)} = \frac{1}{t^2} -
\frac{\omega_0}{3}  - \left( \frac{3 g_2}{40} - \frac{\omega_0
^2}{6} \right) t^2 + \ldots , \quad \xi_{0, 2} ^{(1)} =
\frac{t^3}{5} + \frac{\omega_0}{35} t^5 + \ldots .
\end{equation}

Now, we suppose that $g_{\rm{BF}} = \frac{n (n+1)}{2} \neq 0$.

First, let us consider the condition 1: $a_1 = \frac{4}{n (n+1)},
\, n \in \mathbb{Z}$, i.e., the Lam\'e and Hermite case (i). We
take $n = 1 $ for simplicity, but we keep writing  $g_{\rm{BF}}$
instead 1. In the vicinity $t=0$ we have the  following expansions
for the solutions $\xi_{j, 1} ^{(1)}, \xi_{j, 2} ^{(1)}$ of
(\ref{3.102}) ($n = 1$)
\begin{equation}
\label{3.110} \xi_{j, 1} ^{(1)} = \frac{1}{t} + \frac{B_j}{2} t +
\left( \frac{g_2}{40} - \frac{B_j ^2}{8} \right) t^3 + \ldots,
\quad \xi_{j, 2} ^{(1)} =  \frac{t^2}{3} - \frac{a_j}{30} t^4  +
\ldots ,
\end{equation}
where $B_j = 2 \omega_j - 4 \omega_0 /3$.

There are no logarithms in the expansions around $t=0$ of the
local solutions of the second variational equation
(${\rm{VE}}_2$).

We will show that  a logarithmic term appears in a local solution
of (${\rm{VE}}_3$). For this purpose, it is enough to show that
at least one component of $X^{-1} f_3$ has a nonzero residue at $t
= 0$, see formulae (\ref{2.7}), (\ref{2.8}). We calculate  $j$-th
component $ j = 1, \ldots, N_f$
 of $X^{-1} f_3$, which looks like
\begin{equation}
\label{3.111} (-\xi_{j, 2} ^{(1)} K_j ^{(3)}, \xi_{j, 1} ^{(1)}
K_j ^{(3)} )^T .
\end{equation}

We take
\begin{equation}
\label{3.112} \xi_0 ^{(1)} = \xi_{0, 2} ^{(1)}, \quad \xi_j ^{(1)}
= \xi_{j, 1} ^{(1)} .
\end{equation}
With this choice we find
\begin{equation}
\label{3.113} \xi_{0, 1} ^{(2)} = \frac{1}{t^2} +
\frac{g_{\rm{BF}} N_f}{2}\frac{1}{t} - \frac{\omega_0}{3} +
\ldots, \quad \xi_{0, 2} ^{(2)} = \frac{g_{\rm{BF}}
N_f}{2}\frac{1}{t} + \ldots
\end{equation}
and
\begin{equation}
\label{3.114} \xi_{j, 1} ^{(2)} = \frac{1}{t} + \frac{B_j}{2} t +
\ldots, \quad \xi_{j, 2} ^{(2)} = \frac{t^2}{3} + \ldots .
\end{equation}

Taking the first term in (\ref{3.111}), namely
$$
\mu_3 = -\xi_{j, 2} ^{(1)} K_j ^{(3)} = -\xi_{j, 2} ^{(1)} 2
g_{\rm{BF}} \left[ (\xi_0 ^{(1)})^2 \xi_j ^{(1)} + 2 \bar{q}_0
\left( \xi_0 ^{(1)} \xi_j ^{(2)} + \xi_0 ^{(2)} \xi_j ^{(1)}
\right)  \right]
$$
with the choice (\ref{3.112}) and  $\xi_0 ^{(2)} = \xi_{0,2}
^{(2)}$ and $ \xi_j ^{(2)} = \xi_{j,1} ^{(2)}$ we can see that
$\mu$ has a simple pole at $t = 0$ with residue $-2 g_{\rm{BF}}^2
N_f /3$, which is non-zero. Therefore, the identity component of
the Galois group of (${\rm{VE}}_3$) is not commutative and hence,
in this case, the Hamiltonian system (\ref{1.8}) is not integrable
due to Theorem 3.

Similarly, for $n=2$ and $g_{\rm{BF}}=3$ we have the following
expansions for the solutions $\xi_{j, 1} ^{(1)}, \xi_{j, 2}
^{(1)}$ of (\ref{3.102})
\begin{equation}
\label{3.115} \xi_{j, 1} ^{(1)} = \frac{1}{t^2} - \frac{B_j}{6} +
O (t^2), \quad \xi_{j, 2} ^{(1)} = \frac{t^3}{5} + \frac{B_j
t^5}{70} + \ldots ,
\end{equation}
where $B_j = 4 \omega_0 -2 \omega_j (n=2)$. Let us study first the
expansions of the local solutions of the (${\rm{VE}}_2$) around
$t=0$. The calculation of $\mu_2 = \xi_{j,1} ^{(1)} K_j ^{(2)}$
with $\xi_0 ^{(1)} = \xi_{0,1} ^{(1)}, \xi_j ^{(1)} = \xi_{j,1}
^{(1)}$ gives
$$
\mu_2 = \frac{12}{t^7} - \frac{4 B_j}{t^5} + \frac{\frac{4}{3} B_j
^2 - \frac{12}{5} g_2}{t^3} + \frac{B_j (g_2 - \frac{1}{3} B_j
^2)}{t} + O (t) .
$$
Since $g_2$ depends on $h$, which is arbitrary provided $\Delta =
g_2 ^3 - 27 g_3 ^2 \neq 0$, the only possibility for the residue
of $\mu_2$ to be zero is $B_j = 0$ or $\omega_j = 2 \omega_0$. If
at least exists one $\omega_j$, such that $B_j \neq 0$, then a
logarithm appears in the solutions of (${\rm{VE}}_2$) around
$t=0$.

We proceed with the case when all $B_j = 0$, or equivalently,
$\omega_j = 2 \omega_0, j = 1,\ldots, N_f$. Choosing
\begin{equation}
\label{3.116} \xi_0 ^{(1)} = \xi_{0, 1} ^{(1)} = \frac{1}{t^2} -
\frac{\omega_0}{3} + \ldots, \quad \xi_j ^{(1)} = \xi_{j, 2}
^{(1)} = \frac{t^3}{5}  + \ldots .
\end{equation}
 we find that
$$
\xi_{0, 1}^{(2)} = \frac{1}{t^3} + \frac{1}{t^2} -
\frac{\omega_0}{5 t} - \frac{\omega_0}{3} + \ldots , \qquad
\xi_{0, 2}^{(2)} = \frac{1}{t^3} - \frac{\omega_0}{5 t} + \ldots
$$
and
$$
\xi_{j, 1}^{(2)} = \frac{1}{t^2} + \ldots  , \qquad \xi_{j,
2}^{(2)} = -\frac{3}{5} t^2 + \frac{t^3}{5} + \ldots .
$$
Taking the second term in (\ref{3.111}) $\mu_3 = \xi_{j,1}^{(1)}
K_j ^{(3)}$ with the choice (\ref{3.116}) and $\xi_0 ^{(2)} =
\xi_{0, 2}^{(2)}, \xi_j ^{(2)} = \xi_{j, 2}^{(2)}$ one can see
that $\mu_3$ has a simple pole with a residue $-\omega_0 72/25$,
which is nonzero since $\omega_0 \neq 0$ by assumption.

In either of the cases above, the identity component of the Galois
group of (${\rm{VE}}_2$) or (${\rm{VE}}_3$) is not commutative and
the Hamiltonian system (\ref{1.8}) is not integrable due to
Theorem 3.

Next we consider the case 2.1 with (\ref{3.711}). Here
$n=\frac{1}{2}$ and $g_{\rm{BF}}=\frac{3}{8} $. We take
$$
\xi_0 ^{(1)} = \xi_{0,1} ^{(1)}, \quad \xi_j ^{(1)} = \xi_{j,2}
^{(1)} .
$$
There are no logarithms in the expansions around $t=0$ of the
local solutions of the second variational equation (${\rm{VE}}_2$)
due to (\ref{3.711}). With the above choice, we find that
$$
\xi_0 ^{(2)} = \xi_{0,2} ^{(2)} = \frac{1}{t^3} + O(t), \quad
\xi_j ^{(2)} = \xi_{j,1} ^{(2)} = \frac{1}{\sqrt{t}} - \frac{3}{4}
\sqrt{t} + O( t^{3/2}) .
$$
Then the first term in (\ref{3.111}) has the following expansion
around $t=0$
$$
\mu_3 = -\xi_{j, 2} ^{(1)} K_j ^{(3)} = -\frac{3}{8} \left[
\frac{2}{t^2} - \frac{2 \omega_0}{3 t} + \ldots  \right],
$$
that is, $\mu_3$ has a  pole at $t=0$ with non-zero residue
$\frac{\omega_0}{4}$. Therefore, the identity component of the
Galois group of (${\rm{VE}}_3$) is not abelian and hence, in this
case, the Hamiltonian system (\ref{1.8}) is not integrable due to
Theorem 3.

Finally,  we consider the case 2.3 with (\ref{3.712}). Here
$n=\frac{5}{2}$ and $g_{\rm{BF}}=\frac{35}{8} $. We take
$$
\xi_0 ^{(1)} = \xi_{0,1} ^{(1)}, \quad \xi_j ^{(1)} = \xi_{j,1}
^{(1)} .
$$
There are no logarithms in the expansions around $t=0$ of the
local solutions of the second variational equation (${\rm{VE}}_2$)
due to (\ref{3.712}). With the above choice, we find that
$$
\xi_0 ^{(2)} = \xi_{0,2} ^{(2)} = \frac{5 N_f}{144 t^4} +
\frac{1}{t^3} - \frac{ N_f \omega_0}{224 t^2} - \frac{ \omega_0}{5
t} + O (t^0), $$
$$
\xi_j ^{(2)} = \xi_{j,2} ^{(2)} = t^{7/2} (1 + O (t^2) + t^{-7/2}
\left(\frac{5}{12} - \frac{\omega_j}{99} t^2 + \ldots\right) .
$$
Again the first term in (\ref{3.111}) has the  expansion around
$t=0$
$$
\mu_3 = -\xi_{j, 2} ^{(1)} K_j ^{(3)} = -\frac{175 N_f}{1728 t^4}
- \frac{35}{3 t^3} - \frac{5 N_f \omega_0}{576 t^2} + \frac{7
\omega_0}{12 t} + O (t^0),
$$
that is, $\mu_3$ has a  non-zero residue $\frac{7}{12} \omega_0$
at $t=0$. Therefore, the identity component of the Galois group of
(${\rm{VE}}_3$) is not abelian and hence, in this case, the
Hamiltonian system (\ref{1.8}) is not integrable due to Theorem 3.

{\bf Remark 3}. For arbitrary $n \in \mathbb{Z}$ in $g_{\rm{BF}} =
\frac{n (n+1)}{2}$ one needs to know the exact coefficients in
expansions of the Lam\'e solutions of (\ref{3.102}) and eventually
the expansions of the higher variations. The formulas are quite
involved. However, it is unlikely that the system is integrable
for some $n > 2$.

This finishes the proof of  this part of Theorem 1.

\subsection{The case $C_0 \neq 0, C_1 \neq 0$.}

Here we consider the Hamiltonian (\ref{1.8}) only for two degrees
of freedom (see comments in the next section)
\begin{equation}
\label{5.1} H = \frac{p_0^2}{2} + \omega_0 q_0 ^2 -
\frac{q_0^4}{2} + \frac{C_0^2}{2 q_0^2} + \frac{p_1^2}{2} +
\omega_1 q_1 ^2 + \frac{C_1^2}{2 q_1^2} - g_{\rm{BF}} q_0 ^2 q_1
^2 .
\end{equation}
Denote $\varepsilon := g_{\rm{BF}}$ and assume that $\varepsilon$
is small enough. We can rewrite (\ref{5.1}) as
\begin{equation}
\label{5.2} H = H_0 + \varepsilon H_1 ,
\end{equation}
where
\begin{equation}
\label{5.3} H_0 = \frac{p_0^2}{2} + \omega_0 q_0 ^2 -
\frac{q_0^4}{2} + \frac{C_0^2}{2 q_0^2} + \frac{p_1^2}{2} +
\omega_1 q_1 ^2 + \frac{C_1^2}{2 q_1^2} , \quad H_1 = -  q_0 ^2
q_1 ^2 .
\end{equation}
The unperturbed system ($\varepsilon = 0$) is separable.
\begin{eqnarray}
\label{5.4}
\dot{q}_0 = p_0 , & \dot{p}_0 = - 2 \omega_0 q_0 + 2 q_0 ^3 + \frac{C_0 ^2}{q_0 ^3}, \\
\label{5.5} \dot{q}_1 = p_1 , & \dot{p}_1 = - 2 \omega_1 q_1 +
\frac{C_1 ^2}{q_1 ^3}
\end{eqnarray}
From the proof of Proposition 2 the general solution of
(\ref{5.4}) is found in (\ref{3.222}). From the proof of
Proposition 1 the general solution of (\ref{5.5}) is
\begin{equation}
\label{5.7} q_1 ^2 = \frac{h_1}{2\omega_1} + \sqrt{\frac{C_1 ^2}{2
\omega_1} - \frac{h_1 ^2}{4 \omega_1 ^2} } \sinh 2 i \sqrt{2
\omega_1} (t-t_0), \quad  p_1 = \dot{q}_1 .
\end{equation}

First, we put the Hamiltonian $H$ in the context of the theory
recalled in Section 2. It is assumed that at this point the
variables are real. We introduce action-angle variables $(I,
\varphi)$, so that $H_0 = H_0 (q_0, p_0, I)$. To do so, we need to
find a generating function $S (I, q_1)$ :
$$
(p_1, q_1) \stackrel{S (I, q_1)}{\longrightarrow} (I, \varphi),
\quad p_1 = \frac{\partial S}{\partial q_1}, \quad \varphi =
\frac{\partial S}{\partial I},
$$
such that
\begin{equation}
\label{5.8} \frac{p_1^2}{2} + \omega_1 q_1 ^2 + \frac{C_1^2}{2
q_1^2} = h_1 \to h_1 (I) : = I .
\end{equation}
Note that the real ovals for the curve $(p_1, q_1)$ in (\ref{5.8})
exist for $h_1 > \frac{C_1 ^2 \sqrt{2 \omega_1}}{\sqrt{C_1 ^2}}$.
Then the formula (\ref{5.7}) becomes
\begin{equation}
\label{5.9} q_1 ^2 = \frac{h_1}{2\omega_1} - \sqrt{\frac{h_1 ^2}{4
\omega_1 ^2} - \frac{C_1 ^2}{2 \omega_1}} \sin 2 \sqrt{2 \omega_1}
(t-t_0) .
\end{equation}
The generating function $S$ can be found explicitly, but we do not
need it, we just set
\begin{equation}
\label{5.10} I := \frac{p_1^2}{2} + \omega_1 q_1 ^2 +
\frac{C_1^2}{2 q_1^2}, \quad \varphi : = \int \frac{dq_1}{p_1} .
\end{equation}
 Note that $d I \wedge d \varphi = d p_1 \wedge d q_1$, $\varphi$ is multivalued,
 but $\dot{\varphi} = 1$, that is, $t$ and $\varphi$ are interchangeable.

Next, we fix $I$ to an arbitrary constant greater than $\frac{C_1
^2 \sqrt{2 \omega_1}}{\sqrt{C_1 ^2}}$ and again consider $t, q_0
(t), p_0 (t) $ as complex variables. Our system becomes an
one-and-a-half degrees of freedom system with a Hamiltonian $H =
H_0 + \varepsilon H_1$, where
\begin{equation}
\label{5.11} H_0 =\frac{p_0^2}{2} + \omega_0 q_0 ^2 -
\frac{q_0^4}{2} + \frac{C_0^2}{2 q_0^2} + I, \quad H_1 = - q_0 ^2
\left( \frac{I}{2\omega_1} - \sqrt{\frac{I ^2}{4 \omega_1 ^2} -
\frac{C_1 ^2}{2 \omega_1}} \sin 2 \sqrt{2 \omega_1} (t-t_0)
\right) .
\end{equation}

We need to find a separatrix in the dynamics of $(q_0, p_0)$.
Denote $\tilde{h} = h - I$ and $\tilde{g}_2 =
\frac{16}{3}\omega_0^2 - 4 \tilde{h}$, $\tilde{g}_3 = 4 C_0^2 -
\frac{8}{3}\omega_0 \tilde{h} + \frac{64}{27} \omega_0^3$ (compare
with the corresponding formulas in the Proposition \ref{prop2}).
Let $h^*$ be the biggest real root of
\begin{equation}
\label{5.12} \Delta (\tilde{h}) = \tilde{g}_2 ^3 - 27 \tilde{g}_3
^2 = -64 \left(\tilde{h}^3 -\omega_0 ^2 \tilde{h}^2 -9 C_0 ^2
\omega_0 \tilde{h} + 8 C_0 ^2 \omega_0 ^3 + \frac{27}{4} C_0 ^4
\right) = 0 .
\end{equation}
Assume that $4 \omega_0 ^2 - 3 h^* >0$. Further, we denote
$$
a:= \frac{\sqrt{4 \omega_0 ^2 - 3 h^*}}{3} > 0 .
$$
Then the unperturbed system (\ref{5.11}) has a separatrix
\begin{equation}
\label{5.13} \Gamma_0 : q_0 ^2 (t) = \frac{2}{3} \omega_0 + a +
\frac{3 a}{\sinh ^2 (\sqrt{3 a} t)}, \quad p_0 (t) = \dot{q}_0
(t).
\end{equation}
The perturbed variational equation (PVE) of (\ref{5.11}) along
$\Gamma_{t_0}$ is given by (see \cite{GH,M3})
\begin{equation}
\label{5.14} \frac{d}{d t}
\begin{pmatrix}
\xi \\ \eta \\ \nu
\end{pmatrix} =
\begin{pmatrix}
H_{0, q_0 p_0} & H_{0, p_0 p_0} & H_{1, p_0} \\
-H_{0, q_0 q_0} & -H_{0, q_0 p_0} & -H_{1, q_0} \\
0 & 0 & 0
\end{pmatrix}
\begin{pmatrix}
\xi \\ \eta \\ \nu
\end{pmatrix} ,
\end{equation}
where all coefficients are restricted to $\Gamma_{t_0}$. In order
to study the Galois group of (PVE) we fix the coefficient field
$K$ in (\ref{5.14}). From the expressions for the separatrix
(\ref{5.13}) and the perturbation $H_1$ (\ref{5.11})
$$
K : = \mathbb{C} (e^{\sqrt{3a} t}, e^{2\sqrt{2\omega_1} i t}) .
$$
Then, to obtain the fundamental matrix of (\ref{5.14}) a
quadrature is needed, namely $\delta = \delta (t) = \int
\frac{H_{0, p_0 p_0}}{H_{0, p_0} ^2 }_{|_{\Gamma_0}} d t$ (see
\cite{M3} for details). In our case $\delta   = \int \frac{d
t}{p_0 ^2 (t)}$ equals
\begin{eqnarray*}
  \delta =  \frac{1}{(3a)^3}
( \frac{2\omega_0  +3 a}{12 \sqrt{3a}} \sinh (\sqrt{3a} t) \cosh^3 (\sqrt{3a} t) +  \frac{10\omega_0 + 27a}{8\sqrt{3a}} \sinh (\sqrt{3a} t) \cosh (\sqrt{3a} t) \\
   + \frac{2\omega_0 +12 a}{3\sqrt{3a}} \tanh (\sqrt{3a} t) + \frac{26\omega_0 + 99 a}{8} t ).
\end{eqnarray*}
It is clear that $\delta = \delta (t)$ is uniform and $\delta
\notin K$. Then, the Picard-Vessiot extension of (\ref{5.14})  is
 $L_1 = K (\delta) = \mathbb{C} (e^{\sqrt{3a} t}, e^{2\sqrt{2\omega_1} i t}, t)$. It remains to find $d (t_0)$.
Let $\gamma$ be a loop around the pole $t = 0$. Then simple
calculations give that the Poincar\'e-Arnold-Melnikov integral is
\begin{equation}
\label{5.15} d (t_0) = \int_{\gamma} \{H_0, H_1 \} (q_0 (t), p_0
(t), t - t_0) d t = 12 \pi i a \sqrt{2 \omega_1} \sqrt{\frac{I
^2}{4 \omega_1 ^2} - \frac{C_1 ^2}{2 \omega_1}} \sin 2 \sqrt{2
\omega_1} t_0 .
\end{equation}
It is seen that $d (t_0)$ has simple zeroes and by Theorem 4, the
perturbed separatrix self-intersects transversally. Also since $d
(t_0)$ is not identically zero, the Galois group of the perturbed
variational equation is not abelian \cite{M3}. Hence, when
$\varepsilon = g_{\rm{BF}} \neq 0$ sufficiently small, there is no
additional meromorhic first integral. This finishes the proof of
this part and therefore, the proof of the Theorem 1.

\noindent

$\hfill \blacksquare$

\section{Concluding Remarks}

In this paper we use  variational equations to obtain a necessary
and sufficient condition for integrability of a system which
describes the stationary solutions in the time dependent mean
field equations of Bose--Fermi mixture. Here we make some remarks.

We start with some restrictions to our methods. In subsection 3.1
we don't know how to study the Galois group of a second order
linear equation with quasi-periodic coefficient, that is why we
assume that all $\omega_j$ are equal. It is an open problem
to develop a Picard-Vessiot Theory for the coefficient field $K=
\mathbb{C} (e^{\alpha_1 x}, \ldots, e^{\alpha_m x})$ with
$\alpha_1, \ldots, \alpha_m \in \mathbb{C}$ and to relate this
result with the integrability of the corresponding linear
equation, see \cite{RSh}, p. 408.

In 3.2 we consider $n=1$ and $n=2$ only by technical reasons. It
is highly unlikely that the system is integrable for $n > 2, n \in
\mathbb{Z}$, which is justified by the result in 3.3.  We
notice that the non-integrability result obtained in this case are
also valid for the limiting case $C_0 = 0$ and $C_j = 0, j = 1,
\ldots, N_f$.

In the general case $C_0 \neq 0$ and $C_j \neq 0, j = 1, \ldots,
N_f$ (VE) does not split in nice way as in the previous cases.
Because of this reason, we consider the system (\ref{1.8}) with
two degrees of freedom. Even then, studying the Galois group of
(NVE) is not so simple due to a number of parameters. That is why
we use a perturbational approach, which is still related to the
Differential Galois approach. Furthermore, this approach gives a
dynamical meaning to the algebraic obstructions to integrability.
Note that, the using Poincar\'e-Arnold-Melnikov integrals in more
degrees of freedom for real Hamiltonian systems needs certain
KAM-conditions.

\vspace{2ex}

The above results allow us to think that the system (\ref{1.8}) is
not integrable unless $g_{\rm{BF}} = 0$. Moreover, the formulas
(\ref{4.10101}) and (\ref{3.222}) give the general solution to the
separable system ($g_{\rm{BF}} = 0$).

\vspace{2ex}

{\bf Acknowledgements}
 O. C. acknowledges funding from Bulgarian NSF
Grant DDVU 02/90.

\newpage

\appendix

\section{Necessary conditions for integrability of Hamiltonian systems which have (NVE) of Lam\'e type}

In this appendix we recall some facts concerning the integrability
of Hamiltonian systems with two degrees of freedom, an invariant
plane and which (NVE) are of Lam\'e type. More details can be
found in \cite{MSim,M}. In our case the (NVE) splits into a system
of $N_f$ equations of Lam\'e type, and therefore, these arguments
can be applied.

Classically the Lam\'e equation is written in the form
\begin{equation}
\label{Lame} \ddot{\xi} - [n(n+1) \wp(t) + B] \xi = 0,
\end{equation}
where $\wp (t)$ is the Weierstrass function with invariants $g_2$
and $g_3$, satisfying $\dot{v}^2 = 4 v^3 - g_2 v - g_3$ with
$\Delta = g_2 ^3 - 27 g_3 ^2 \neq 0$.

The known (mutually exclusive) cases of closed form solutions of
(\ref{Lame}) are:

(i) The Lam\'e and Hermite solutions. In this case $n \in
\mathbb{Z}$ and $g_2, g_3, B$ are arbitrary parameters;

(ii) The Brioschi-Halphen-Crowford solutions. Here $m:= n + 1/2
\in \mathbb{N}$ and the parameters $g_2, g_3, B$ must satisfy an
algebraic equation.

(iii) The Baldassarri solutions. Now $n + 1/2 \in \frac{1}{3}
\mathbb{Z} \cup \frac{1}{4} \mathbb{Z} \cup \frac{1}{5} \mathbb{Z}
\setminus \mathbb{Z}$ with additional algebraic relations between
the other parameters.

Note that in the case (i) the identity component of the Galois
group $G^0$ is of the form $\begin{pmatrix}
1 & 0 \\
\nu & 1
\end{pmatrix}$
 and in the cases (ii) and (iii) $G^0 = id$ ($G$ is finite).
And these are the all cases when the Lam\'e equation is
integrable.

Now consider a natural two degrees of freedom Hamiltonian
\begin{equation}
\label{ham2} H = \frac{1}{2} (p_1 ^2 + p_2 ^2) + V (q_1, q_2),
\end{equation}
$q_j (t) \in \mathbb{C}, p_j (t) = \dot{q}_j, j=1, 2$. We assume
that there exists a family of solutions of the form
$$
\Gamma_h : q_2 = p_2 = 0, \quad q_1 = q_1 (t, h), \quad  p_1 (t,
h) = \dot{q}_1 (t, h)
$$
and $q_1 (t, h)$ is a solution of
$$
\frac{1}{2} \dot{q}_1 ^2 + \varphi (q_1) = h, \quad h \in
\mathbb{R} .
$$
The (NVE) along $\Gamma_h$ is
\begin{equation}
\label{nve} \ddot{\xi} - \alpha (t, h) \xi = 0 ,
\end{equation}
where $\alpha (t, h) = \alpha (q_1 (t, h))$ is such that
(\ref{nve}) is of type (\ref{Lame}).

In \cite{MSim,M} the type of the potentials $V$ with this property
are obtained as well as the necessary conditions for the
integrability of the Hamiltonian systems with the Hamiltonian
(\ref{ham2}). In order to formulate the result we need certain
additional quantities.

Since $\alpha (t, h)$ depends linearly on $\wp (t)$, then
$\dot{\alpha}^2$ is a cubic polynomial in $\alpha$, depending also
in $h$, namely
\begin{equation}
\label{apl} \dot{\alpha}^2 : = P (\alpha, h) = P_1 (\alpha) + h
P_2 (\alpha) .
\end{equation}
The following coefficients are introduced
\begin{equation}
\label{coeff} P (\alpha, h) = (a_1 + h a_2) \alpha^3 + (b_1 + h
b_2) \alpha^2 + (c_1 + h c_2) \alpha + (d_1 + h d_2) .
\end{equation}
Now we are ready to give the corresponding result. Note that the
following Theorem gives necessary conditions only from the
analysis of the first variational equation.
\begin{thm}
\label{thA} (Theorem 6.2 \cite{M}). Assume that a natural
Hamiltonian system has (NVE) of Lam\'e type, associated to the
family of solutions $\Gamma_h$, lying on the plane $q_2 = 0$ and
parametrized by the energy $h$. Then, a necessary conditions for
integrability is that the related polynomials $P_1$ and $P_2$
satisfy $a_2 = 0$, and one of the following conditions holds:

\vspace{2ex}

\noindent 1. $a_1 = \frac{4}{n (n+1)}$ for some $n \in
\mathbb{N}$;

\vspace{2ex}

\noindent 2. $a_1 = \frac{16}{4 m^2 - 1}$ for some $m \in
\mathbb{N}$. Then, assuming the conjecture above is true, one
should have $b_2 = 0$ and we should be in one of the following
cases:

\vspace{1ex}

2.1) $m = 1$ and $b_1 = 0$,

\vspace{1ex}

2.2) $m = 2$ and $c_2 = 0, \, 16 a_1 c_1 + 3 b_1 ^2 = 0$,

\vspace{1ex}

2.3) $m = 3$ and $16 a_1 d_2 + 11 b_1 c_2 = 0, \, 1024 a_1 ^2 d_1
+ 704 a_1 b_1 c_1 + 45 b_1 ^3 = 0$,

\vspace{1ex}

2.m) $m > 3$. Then, we should have $b_1 = 0$ and, furthermore,
either $c_1 = c_2 = 0$ if $m$ is congruent with $1, 2, 4$ or $5$
modulo $6$, or $d_1 = d_2 = 0$ if $m$ is odd;

\vspace{2ex}

\noindent 3.  $a_1 = \frac{4}{n (n+1)}$ with $n + 1/2 \in
\frac{1}{3} \mathbb{Z} \cup \frac{1}{4} \mathbb{Z} \cup
\frac{1}{5} \mathbb{Z} \setminus \mathbb{Z}$, $b_2 = 0$ and either
$c_2 = 0, b_1 ^2 - 3 a_1 c_1 = 0$ or $c_2 b_1 - 3 a_1 d_2 = 0, 2
b_1 ^3  - 9 a_1 b_1 c_1 + 27 a_1 ^2 d_1 = 0$.
\end{thm}
It is clear that the condition 1. in the above Theorem gives the
Lam\'e and Hermite solutions (i), the condition 2.-- the
Brioschi-Halphen-Crowford solutions (ii), and the condition 3. --
the Baldassarri solutions (iii).

\end{document}